\documentclass[superscriptaddress,pre,reprint,showpacs,twocolumn]{revtex4-1}

\usepackage[utf8]{inputenc}
\usepackage{amsmath, amsfonts}
\usepackage{graphicx}
\usepackage{natbib}
\usepackage{caption}
\usepackage{subcaption}

\usepackage{hyperref}

\usepackage{mathptmx}

\newcommand{\defeq}{:=}
\newcommand{\mean}[1]{\left \langle #1 \right \rangle}
\newcommand{\rd}[1]{\mathrm{d}{#1} \,}
\newcommand{\RR}{\mathbb{R}}

\newcommand{\indicatorsymbol}{\mathbf{1}}
\newcommand{\indicator}[1]{\indicatorsymbol_{ \{   #1 \} } } 
\newcommand{\etal}{et al.\ } 

\begin{document}

\title{Exact hopping and collision times for two hard discs in a box}

\author{W.~P.~Karel Zapfe}
\email{karelz@cinvestav.mx}
\affiliation{Departamento de Física, Facultad de Ciencias, Universidad Nacional Autónoma de México, Ciudad Universitaria, Ciudad de México 04510, Mexico}

\affiliation{Departamento de Farmacobiología, CINVESTAV, IPN, Calzada de los Tenorios \#235, Col.~Granjas Coapa, Ciudad de México 14330, Mexico}

\author{David P.~Sanders}
\email{dpsanders@ciencias.unam.mx}
\affiliation{Departamento de Física, Facultad de Ciencias, Universidad Nacional Autónoma de México, Ciudad Universitaria, Ciudad de México 04510, Mexico}

\author{Rosa Rodríguez-Mota}
\email{rosarm@physics.mcgill.ca}
\affiliation{Department of Physics and the Centre for Physics of Materials, McGill University, Montreal, Quebec, Canada H3A 2T8}

\begin{abstract}
We study the molecular dynamics of two discs undergoing Newtonian (``inertial'') dynamics, with elastic collisions in a rectangular box.
Using a mapping to a billiard model and a key result from ergodic theory, we obtain exact, analytical expressions for the mean times between the following events:
hops, i.e.~horizontal or vertical interchanges of the particles; wall collisions; and disc collisions.
To do so, we calculate
volumes and cross-sectional areas in the four-dimensional configuration space.
We compare the analytical results against Monte Carlo and molecular dynamics simulations, with excellent agreement.
\end{abstract}

\maketitle

\section{Introduction}


Collision times play a fundamental role in statistical mechanics, since properties
such as mixing, reaction and diffusion rates depend on them \cite{Boltz72, Tolman, VanKampen}.
A paradigmatic model, introduced by Boltzmann almost 150 years ago \cite{Boltz72, SzaszBook00},
consists of a fluid of hard spheres colliding with one other, either in a periodic torus or in
a rectangular box.

The simplest such system that is non-trivial is that of two discs in a box.
The discs move rectilinearly until they undergo
elastic collisions, either with a wall of the box, or with one another; this has been called ``inertial motion'', to distinguish it from Brownian motion \cite{Bowles04}.
Phenomena such as transport, collision and reactions
have been studied in such systems, both analytically 
 \cite{Awazu01, Munakata02, Suh05} and numerically \cite{MacElroy2004, MacElroy2005}.
The events of physical interest in this system are \emph{collisions} of the particles, with the box or with one other, and \emph{hopping},
in which two particles interchange position, either vertically or horizontally,
and; this plays an important role in the dynamics of confined fluids, for example \cite{Bowles04}.

Previous work
by Bowles \etal gives results for hopping
for both inertial and Brownian dynamics \cite{Bowles04}, defined as 
the mean \emph{first-passage} time, using arguments from transition state theory 
to provide expressions for the general behavior of these quantities, as a function of the
disc radii. 
Further work has expanded along that line  \cite{Suh05, Ball09}.
There is also a statistical thermodynamic treatment by Munakata \etal, 
who study the partition function, pressure,
and temperature in the system \cite{Munakata02, Munakata06}.

In this paper we obtain \emph{exact} analytical expressions, not for the first-passage time,
but rather for mean inter-hop and inter-collision times of two discs under 
inertial (Newtonian) motion, as a function of the geometrical parameters of the system. Such a treatment is possible
 since the dynamics of $N$ hard discs in $d$ spatial dimensions
 is equivalent to a \emph{billiard model} in the $Nd$-dimensional configuration space, i.e. a \emph{single} point particle colliding elastically 
with suitable objects (``scatterers'') \cite{SzaszBook00}. 
In particular, the system of two discs in a box is hyperbolic (chaotic) and ergodic when the discs are able to pass each other \cite{Sim99}. Otherwise, the system decomposes into
two or four disjoint ergodic components, on each of which the dynamics is still hyperbolic.

We can thus apply a key result from ergodic theory on mean collision times \cite{Chernov97} to obtain analytical expressions 
for times between hops, between wall collisions, and between disc collisions, in terms of volumes and areas in configuration space. We verify our results by comparing the geometrical calculations with rejection-sampling Monte Carlo simulations, and the dynamical results with molecular dynamics simulations.
In certain asymptotic regimes, we recover previously obtained power-law exponents \cite{Bowles04}.

The main difficulty in the analysis is correctly accounting for several geometric factors with different origins that occur in both the
analytical and numerical calculations. All numerical code, in the Julia language \cite{Julia}, is available online \cite{repo}.


\section{Model: Two hard discs in a rectangular box}

\begin{figure}[h]
  \begin{center}
    \includegraphics[width=0.40\textwidth]{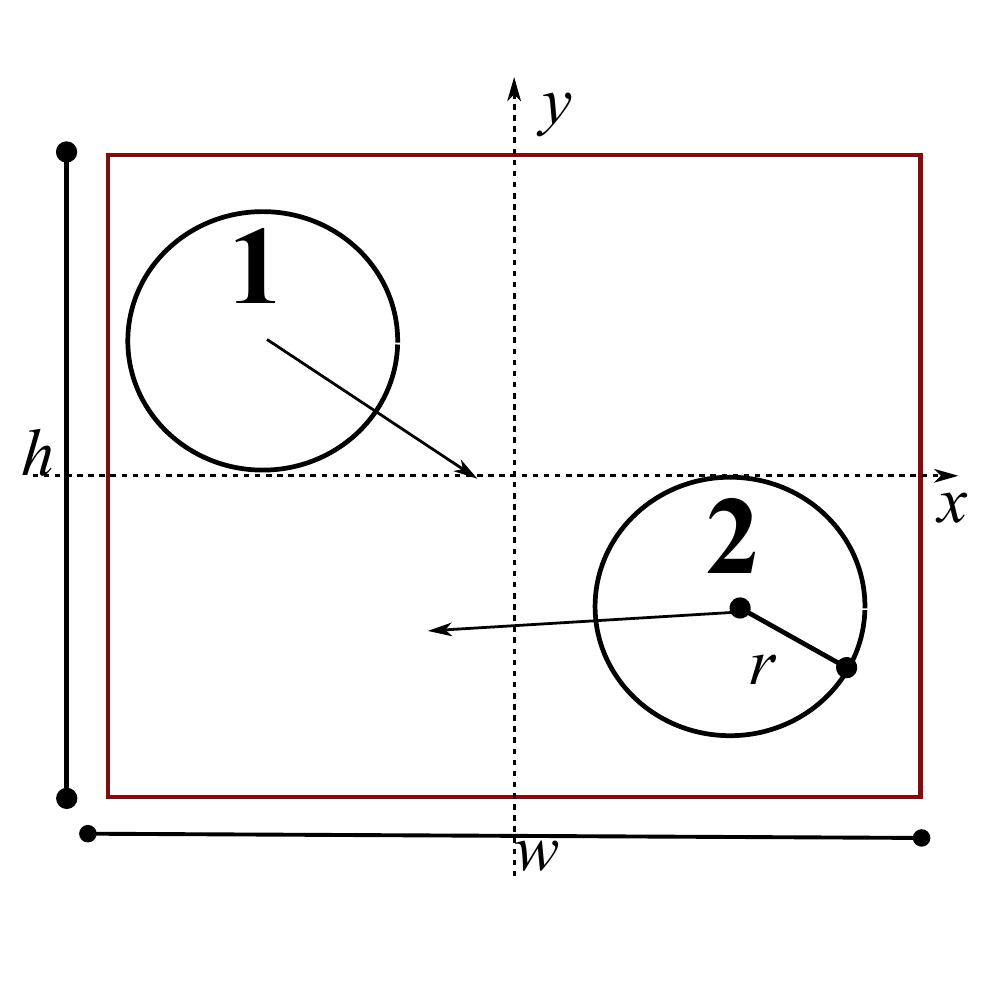}
  \end{center}
  \caption{The billiard and its parameters. Coordinates
    have their origin at the geometrical center of the 
    billiard table.}\label{billar01}
\end{figure}

We consider two hard discs with equal radius $r$,
moving inside a box of width $w$ and height $h$; see Fig.~\ref{billar01}. 
The discs move inertially in the absence of forces, 
following straight line trajectories,
and undergo elastic collisions with one 
other and with the walls of the box.

We denote the position of the center of the $i$th disc by 
$\mathbf{x}_i = (x_i, y_i)$, and its velocity by $\mathbf{v}_i = (u_i, v_i)$ for $i=1,2$. Since the discs are hard, 
their centers are restricted to the region 
$(x_i, y_i) \in [-a,a] \times [-b, b]$, where 
$a \defeq a(r) \defeq \frac{w}{2} - r $ and
$b \defeq b(r) \defeq \frac{h}{2} - r $.

The exclusion condition preventing the discs from overlapping is $(x_1-x_2)^2 + (y_1-y_2)^2 \ge (2r)^2$.
It is thus useful to perform an orthogonal transformation (rotation) to the following new coordinates:
\begin{equation}\label{cambiocoor01}
  \begin{split}
 x  \defeq \frac{x_1 - x_2}{\sqrt{2}};  &
\quad X  \defeq \frac{x_1 + x_2}{\sqrt{2}};  \\
 y  \defeq \frac{y_1 - y_2}{\sqrt{2}}; & 
\quad Y  \defeq \frac{y_1 + y_2}{\sqrt{2}}.
\end{split}
\end{equation}

In these coordinates, the configuration space is given by the following
intervals:
$x \in [-a \sqrt{2}, +a \sqrt{2}]$ with 
$X \in [-a \sqrt{2} + |x|, a \sqrt{2} - |x|]$; and similarly for $y$ and $Y$, replacing $a$ by $b$.
The non-overlapping constraint then becomes $x^2 + y^2 \ge 2 r^2$.
The horizontal and vertical coordinates transform independently
from one other, and the Jacobian of this transformation is  equal to $1$.

These constraints define a four-dimensional
rectangular prism, in which is embedded an excluded cylinder with a three-dimensional surface
(codimension 1).
This cylinder has radius $r\sqrt{2}$ and lies
on a diagonal between the $X$ and $Y$ axes.
The prism surface is the outer boundary of the configuration space,
while the cylinder is an excluded volume, the surface of which
acts as a reflecting inner boundary.
The dynamics of the two discs is equivalent to a billiard model in this 4-dimensional space, in which 
a point particle undergoes free flight until
it hits a wall, where it undergoes an elastic reflection.
The outer boundaries are flat, so the
hyperbolicity is due to the inner semi-dispersing
boundary \cite{Sim99}, which corresponds to the collision of
the two discs.

We take the mass of each disk as $m=1$, so that the kinetic energy
is $\frac{1}{2}(\mathbf{v}_1^2 + \mathbf{v}_2^2)$. We restrict attention to the energy surface with
$E = \frac{1}{2}$, so that the disc velocities satisfy $\mathbf{v}_1^2 + \mathbf{v}_2^2 = 1$.
Other values of the mass or energy correspond to a simple rescaling of the dynamics, with velocities differing
by a factor of
$\sqrt{2E/m}$, and corresponding factors in the times to be determined below.

\subsection{Hopping dynamics}
\begin{figure}[h]
  \begin{center}
    \includegraphics[width=0.45\textwidth]{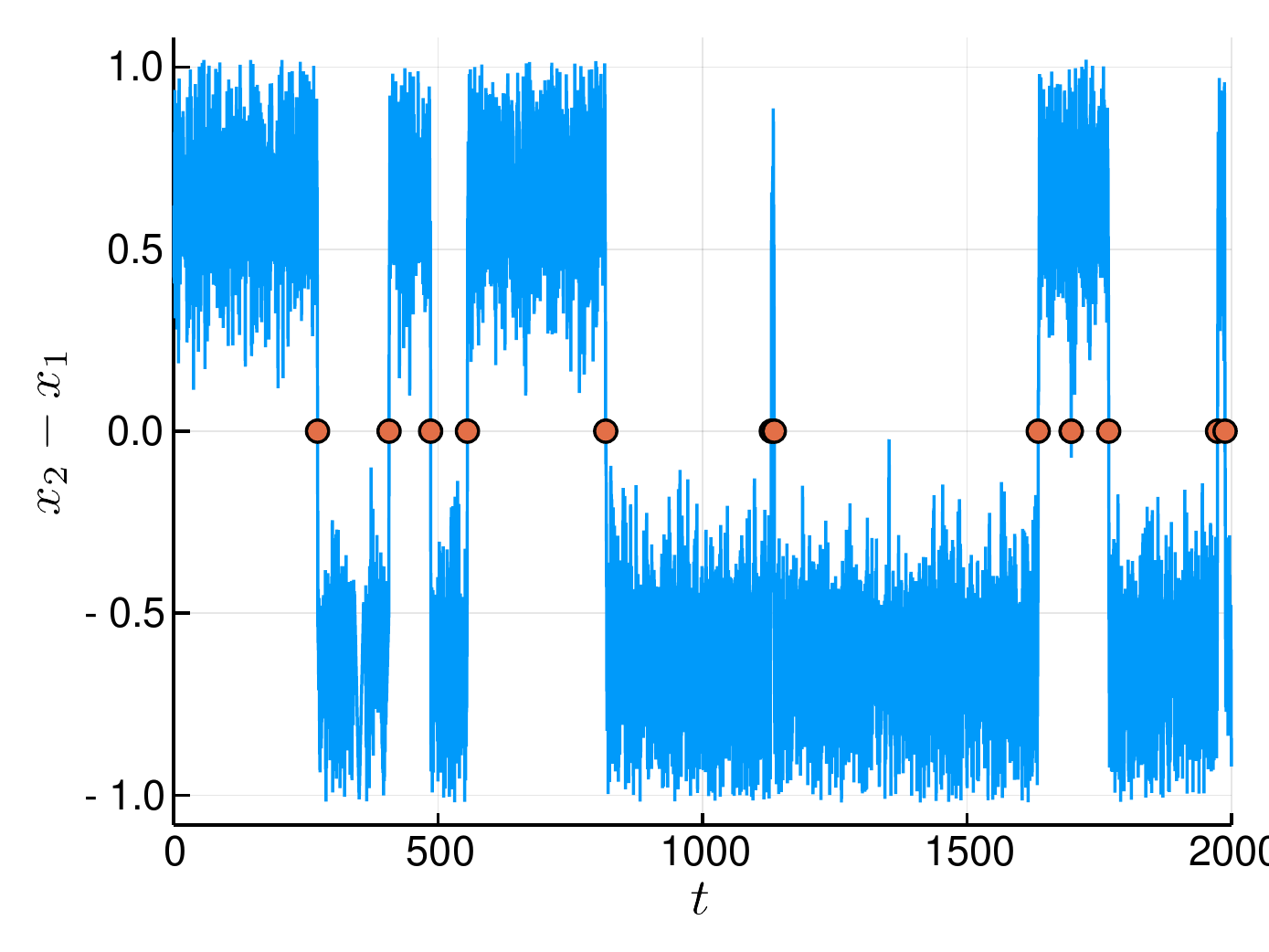}
  \end{center}
  \caption{Hopping dynamics: $x_2 - x_1$ as a function of time $t$. The points show the hopping times. Parameters:  $w = 1.5$, $h = 1.0$, $r = 0.24$.
  \label{fig:hopping-dynamics}  
    }
\end{figure}

An example of hopping dynamics is shown in Fig.~\ref{fig:hopping-dynamics}, for parameter values for which the two discs are trapped for long periods without
being able to undergo horizontal hops. This was simulated using the algorithm described below. We plot the difference $x_2 - x_1$ as a function of time $t$; horizontal hops correspond to the moments at 
which this quantity crosses through $0$.  

We see a bistable-type behavior corresponding to the two possible horizontal arrangements of the two discs. 
The goal of this paper is to calculate the mean time between adjacent hops. Note that, as seen in the figure, these statistics include pairs (and possibly longer sequences) of events that occur very close to each other in time, corresponding to consecutive hops without intermediate randomization. Although we do not necessarily wish to count these as true hopping events, it is not clear how to exclude them.

\section{Mean collision time for billiard models}

\label{knownfacts}

A system of $N$ hard spheres confined by hard walls in a $d$-dimensional
space may be treated as a billiard system 
in which a single point particle undergoes free motion between reflecting obstacles 
in a $ (d N) $-dimensional configuration space \cite{Sinai70, Sim99, MarkChern}. 
This can be thought of as a mean return time to the $(d-1)$-dimensional 
(i.e. co-dimension $1$) cross-section given by the wall boundaries.

For a particle with unit velocity moving in a general billiard table, there is 
an exact expression for the mean return time between consecutive
intersections of the particle with a given surface \cite{Chernov97}:

\begin{equation}\label{meanfreetime}
 \mean{\tau} = \frac{|Q|}{|A|} \cdot \frac{|S^{d-1}|} {|B^{d-1}|} \cdot \frac{1}{s}.
\end{equation}

Here, $|Q|$ denotes the $d$-dimensional volume of the available 
space in the billiard;
 $|S^{d-1}|$ is the $(d-1)$-dimensional area of the unit sphere in $\RR^d$, given by
\begin{equation}
  |S^{d-1}| = \frac{2 \pi^{d/2}}{\Gamma(d/2)},
\end{equation}
where $\Gamma(\cdot)$ is the gamma function; and 
$|B^{d-1}|$ is the volume of the unit ball 
in $\RR^{d-1}$, given by $|B^{d-1}| = |S^{d-2}| / (d-1)$.
$|A|$ denotes the $(d-1)$-dimensional cross-sectional area of the surface in question.
The extra factor of $s$ is a ``sidedness'' factor, explained below.
Note that Machta and Zwanzig \cite{MachtaZwan} used a similar method to derive an escape 
time across a virtual boundary by treating it as a recurrence time.

In our case, we are interested in the mean return time to 
several types of co-dimension-$1$ cross section.
The first, giving the ``hopping time'', 
is defined by the moment
at which the two discs interchange their horizontal or vertical position, i.e.
when they cross one of the two surfaces
\begin{equation} \label{condchoque}
x_1 = x_2  \qquad \text{or} \qquad y_1 = y_2.
\end{equation}
At such an event, the cross-section can be hit from \emph{either} side,
e.g.~disk 1 can be traveling from left to right or the other way around
at the moment when the interchange of positions take place in a horizontal hop.
The area $A$ of the surface is, then, 
effectively twice as large, or equivalently the mean return time to the surface is reduced by half; this is taken into account by taking the ``sidedness'' factor $s=2$ in the formula for $\mean{\tau}$.

The other events of interest correspond to collisions of a specific
disc with a specific wall, and the mutual collision between the two discs.
In both of these cases, the collision surface can be reached only from one side in configuration space, corresponding to $s = 1$.

To calculate the mean times of interest, it is thus necessary to calculate
the 4-dimensional volume $V$ of the available space, and the 3-dimensional cross-sectional area $A$ 
for each event of interest. 

\section{Calculation of volume and areas}

In this section, we calculate analytically the available volume and the cross-sectional areas required.

\subsection{Volume of available space}

\begin{figure}[h]
  \begin{center}
    \includegraphics[width=0.45\textwidth]{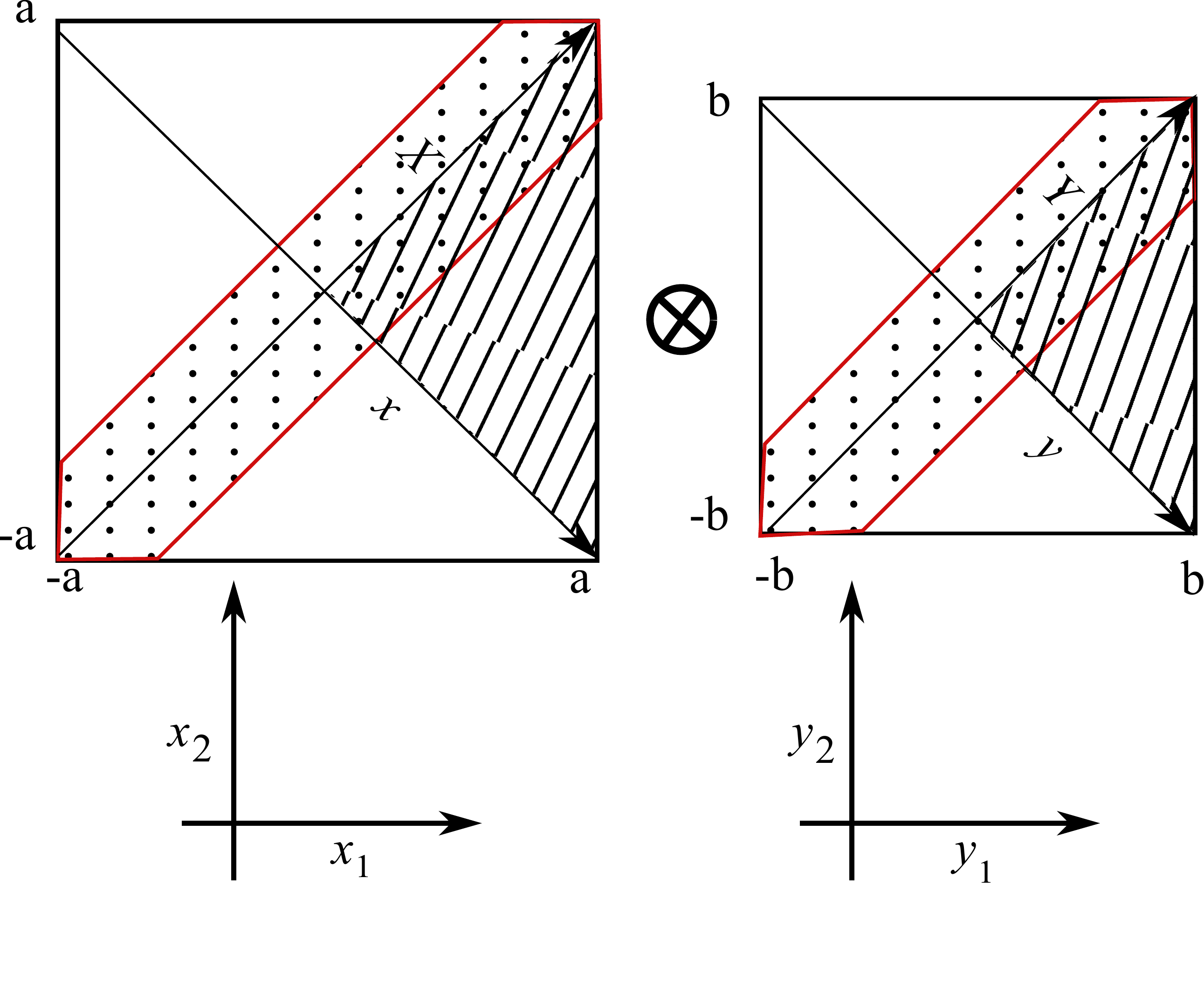}
  \end{center}
  \caption{The space to integrate is the product of the spaces
    corresponding to the horizontal and vertical coordinates. The dotted
    bands show the excluded set, that is, the complement of $Z$, where the condition 
    $ (x_1-x_2)^2 + (y_1-y_2)^2 \ge (2r)^2 $ is not satisfied.
    The width of each band is dependent on the particular 
    point of evaluation
    on the other subspace, which ranges from 0 to $r\sqrt{2}$. Due to symmetry, 
    we evaluate the integral for $X,Y,x,y > 0$ (indicated with hatching in the
    figure), and multiply by 16.
    \label{diagintegral01}  
    }
\end{figure}

We denote by $Z := \{ \mathbf{x} \in \mathbb{R}^4: (x_1-x_2)^2 + (y_1-y_2)^2 \ge (2r)^2 \}$
the complement of the cylinder in configuration space, where $\mathbf{x} := (x_1, x_2, y_1, y_2)$.
The four-dimensional available volume $V_\text{free}$ is then given by
\begin{align}\label{volindic}
V_\text{free} &= 
\iiiint
\limits_{\substack{x_1, x_2 = -a \\ y_1, y_2 = -b}}^{\substack{x_1, x_2 = a \\ y_1, y_2 = b}}
\rd x_1 \rd x_2 \rd y_1 \rd y_2 
\, \indicator{ (x_1-x_2)^2 + (y_1-y_2)^2 \ge (2r)^2 } \\
&=
\int  \mathrm{d} \mathbf{x} \, \indicatorsymbol_Z(\mathbf{x});
\end{align}
here, $\int  \mathrm{d} \mathbf{x}$
denotes a four-dimensional integral over the whole volume, and 
$\indicatorsymbol_Z$ is the indicator function of the set $Z$, 
given by $\indicatorsymbol_Z (\mathbf{x}) = 1$ if $\mathbf{x} \in Z$, and $=0$ if $\mathbf{x} \notin Z$.
Fig.~\ref{diagintegral01} shows a diagram of the product of
spaces that give rise to the whole configuration space.
Recall that the dimensions $a$ and $b$ of the available configuration space are functions of the disc radius $r$, 
but we suppress this explicit dependence for simplicity.

It is easiest to represent the  
excluded cylinder in the coordinates defined in 
Eq.~\eqref{cambiocoor01}:

\begin{equation}\label{integraltotal}
  V_\text{free} = 
  \iint\limits_{\substack{ x=-a \sqrt{2} \\ y=-b \sqrt{2}}}
  ^{\substack{ x=a \sqrt{2} \\ y=b \sqrt{2}}}
   \mkern-9mu \rd x \rd y 
  \iint \limits_{\substack{X=-a\sqrt{2} + |x| \\ Y=-b \sqrt{2} + |y| }}^
  {\substack{X=a\sqrt{2} - |x| \\ Y=b \sqrt{2} - |y| }}
   \mkern-18mu  \rd X \rd Y
\, \indicator{ x^2 + y^2 \ge 2r^2  }.
\end{equation}

Since $X$ and $Y$ do not appear in the integrand, the corresponding integrals are trivial, giving
\begin{align}
  V_\text{free}  &=
  \iint\limits_{\substack{ x=-a \sqrt{2} \\ y=-b \sqrt{2}}}
  ^{\substack{ x=a \sqrt{2} \\ y=b \sqrt{2}}}
   \mkern-9mu \rd x \rd y 
2 \left( a \sqrt{2} - |x| \right) \, 2 \left( b \sqrt{2} - |y| \right) \indicator{ x^2 + y^2 \ge 2r^2 } \\
&=
16  \iint\limits_{\substack{ x= 0 \\ y=0 }}
  ^{\substack{ x=a \sqrt{2} \\ y=b \sqrt{2}}}
   \mkern-9mu \rd x \rd y  
\left( a \sqrt{2} - x \right) \left( b \sqrt{2} - y \right) \indicator{ x^2 + y^2 \ge 2r^2 },
\end{align}
where we have used the symmetry visible in Fig.~\ref{diagintegral01}.
Thus $V_\text{free} = 16(I_1 + I_2)$, where $I_1$ is the region where the range of
values of $y$
is affected by the exclusion condition, and $I_2$ is where the exclusion condition has
no effect on that range.
 In figure \ref{diagintegral01} this would correspond to the dotted and
white areas of the left diagram; see also Fig.~\ref{CasosIntegra} in the appendix. 
We have
\begin{align}
  I_1 &= \iint\limits_{\substack{x=0  \\ y = \sqrt{ 2r^2 - x^2}}}
    ^{\substack{x=r\sqrt{2}\\ y=b \sqrt{2}}} \! \rd x \rd y
\left( b \sqrt{2} - y \right)  \left( a \sqrt{2} - x \right) \\
&= 	
2 a b^{2} r  + \textstyle \frac{1}{6} (a+b) (2r)^{3} - \frac{1}{32}  (2r)^{4} - \frac{1}{4} {\left(\pi a b + b^{2}\right)} (2r)^2,
\end{align}
and
\begin{align}
  I_2 &= \iint\limits_{\substack{x=r\sqrt{2}  \\ y=0 }}
    ^{\substack{x=a\sqrt{2}\\ y=b \sqrt{2}}} \! \rd x \rd y
  \left( b \sqrt{2} - y \right)  \left( a \sqrt{2} - x \right)  \\
&=	
{\left( a^{2} - 2ar +   r^{2}\right)} b^{2}.
\end{align}
Thus 
\begin{align}\label{volumeabd}
 V_\text{free}
 & =  16 a^{2} b^{2}  - 16 \pi a b r^{2} + \textstyle \frac{64}{3} (a+b) r^{3}  - 8 r^{4} \\
&=: V_\text{prism} - V_\text{cyl},
\end{align}
as previously obtained by Munakata and Hu \cite{Munakata02}.
It is useful to split this expression into the total volume of the prism, 
$V_\text{prism}=16 a^2 b^2$, and the cylindrical volume excluded by the overlapping
condition, 
 $V_\text{cyl} =  16 \pi a b r^{2} - \textstyle \frac{64}{3} (a+b) r^{3}  + 8 r^{4}$.

The substitutions $a(r)\leftarrow (w-2r)/2$ and $b(r)\leftarrow (h-2r)/2$ give us
 the volume as a function of the radius, for fixed table size:
\begin{equation}\label{volumewhd}
 V_\text{free} 
= (w-2r)^{2} (h-2r)^{2}  - 
 \pi (w-2r)(h-2r) 4 r^{2} + 
\textstyle \frac{32}{3} (w+h-2r) r^{3}  
- 8^{4}.
\end{equation}

Note that the above formula is correct only when both
vertical and horizontal hopping are possible, i.e., when $h, w > 4r$.
If this is not the case, the 
integration limits for $X$ and $Y$ in \eqref{integraltotal} are altered; 
see Appendix~\ref{app:area_volume} for the corresponding results.

As an example, consider the case $w \geq h$, in which vertical hopping is possible
but horizontal hopping is not.
There are two sub-cases: if
$ h \leq  w < 2 h $, there is a value of $r$ above which hopping is no longer possible,
but the discs still fit in the table. For $2 h \leq w $, vertical hopping is
possible only up to $r = h/2$.
The configuration space splits into disjoint components, but 
thanks to the symmetry of
the problem, in some cases the cross-section areas and 
volumes become disjoint components sharing the same fraction of
the total volume, making the transition continuous. In other cases, we instead end up with 
a discontinuity in the formulas,
corresponding to a factor of 2 or 4.

For example, when $h/4 < r < w/4$ 
defining 
$c = \sqrt{4r^2-b^2}$, we have
\begin{multline}\label{VolumenCasoFeo}
V_{h/4<r<w/4} = 32abr^2 \left[ \arccos(b/2r)-\arccos(a/2r) \right]\\
+\frac{64 r^3}{3 } \left[ a((b-a)/2r)-b(c/2r+\sqrt{4r^2-a^2}/2r) \right]\\
+16 \left[ a b^2 c (4\sqrt{2}-1-\sqrt{2}/3) 
  +c^2b^2 (\sqrt{2}/3-1) \right]\\
-2r^2 (b^2-a^2)
\end{multline}

When $r$ is larger than both $h/4$ and $w/4$, one must take
into account a similar
contribution which inverts the roles of $a$ and $b$.

\begin{figure}[h]
\centering
\includegraphics[width=0.45\textwidth]{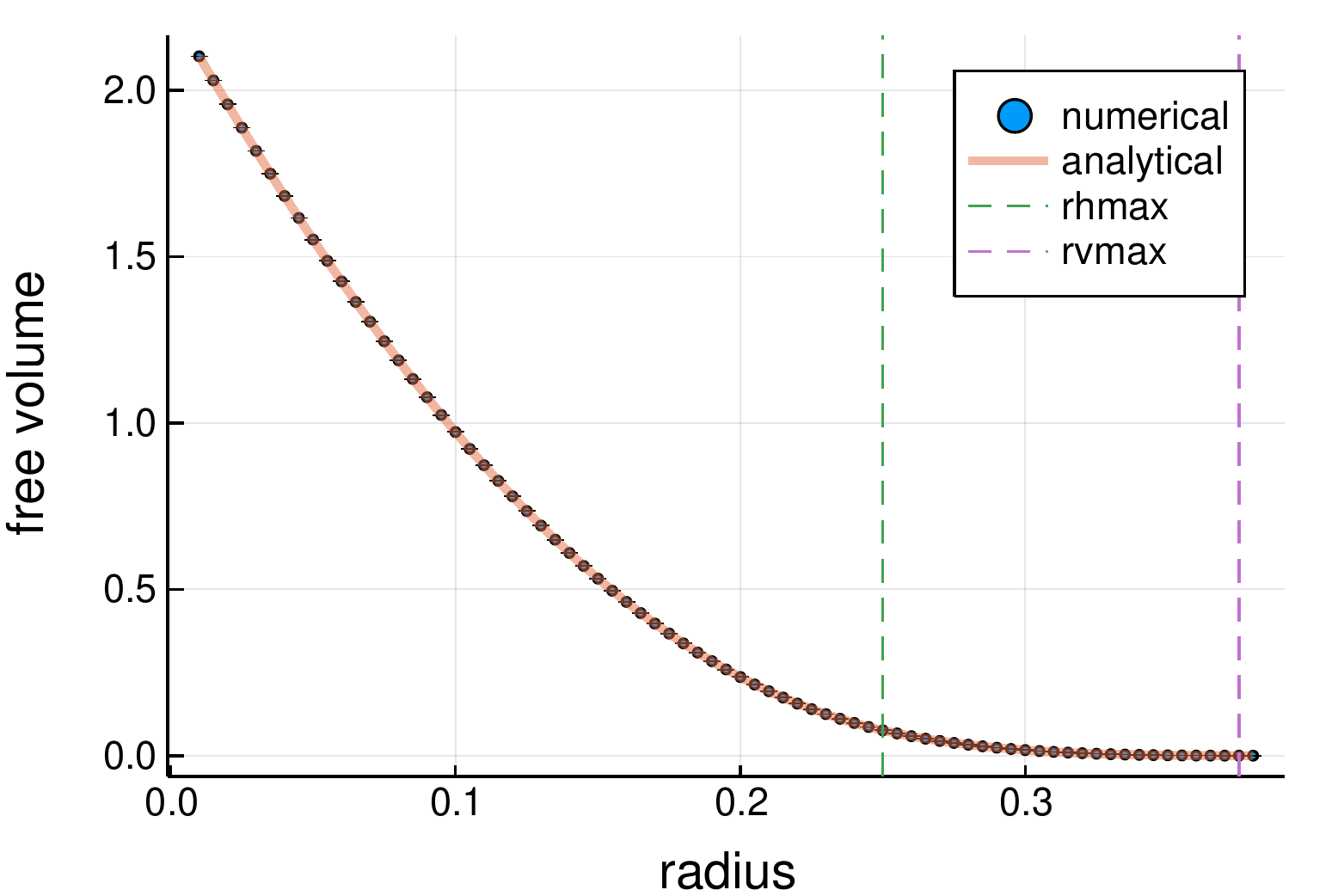}
\caption{Available volume in configuration space as a function of radius. The
  line labeled $rhmax$ is the limit for horizontal hopping, and the one labeled
  $rvmax$ is the limit for vertical hopping. The transition between the formulas
is continuous.}
\label{VolMonteC}
\end{figure}

We have verified our expressions for the available volume numerically using 
standard rejection-sampling Monte Carlo simulations: we 
generate uniform random positions for the disc centers in 
$[-a,a] \times [-b,b]$ and 
count the fraction of initial conditions for 
which the two discs do not overlap (rejecting those where is overlap).
The results are shown in Fig.~\ref{VolMonteC}.
We chose the values $w=1.5, h=1$ for the numerical integration since all the different
cases of the volume formulas occur for these values: all hops possible, only vertical
hops possible, no hops possible. 

In the following, for simplicity we shall give results only for the case $r < \text{min}(\frac{w}{4}, \frac{h}{4} )$.


\subsection{Cross-sectional areas}\label{areahop}

The terms of the form $|A|$ in \eqref{meanfreetime} are surface areas of 3-dimensional surfaces (manifolds) $S$ embedded in  configuration space,
defined by algebraic equations of the form $g(\mathbf{x}) = 0$, so that $S = g^{-1}(0)$ is the zero set of $g$.
The surface area of $S$ is then given by
\begin{equation}
A(S) = \int \indicatorsymbol_Z(\mathbf{x}) \, \delta(g(\mathbf{x})) \, \mathrm{d} \mathbf{x}.
\label{eq:surface-area}
\end{equation}
To evaluate this, we use the following coarea formula
\cite[section 6.1]{Hormander83} 
\begin{equation}
\int_{\mathbf{R}^d} f(\mathbf{x}) \, \delta(g(\mathbf{x})) \, \mathrm{d} \mathbf{x} = \int_{g^{-1}(0)}\frac{f(\mathbf{x})}{\| \mathbf{\nabla}g(\mathbf{x}) \|} \, \mathrm{d}S,
\label{eq:surface-dirac}
\end{equation}
where the integral on the right-hand side is over the surface $g^{-1}(0)$.
\cite{Zappa2018}

The appearance of the normalization factor $\| \mathbf{\nabla}g(\mathbf{x}) \|$ can be understood intuitively by considering how to verify numerically these surface areas. One possible technique is to use rejection sampling to sample the volume given by $|g(\mathbf{x})| \le \eta$ for some small value of $\eta$. Points will be accepted or rejected according to their orthogonal distance to the surface, which is the factor in the denominator in the above expression. Our case is simple in this respect: most of our surfaces are either parallel to the coordinates or are in some way ``diagonal'', giving constant factors.



\subsubsection{Hopping}

Vertical hopping occurs when the vertical positions of the two discs are equal: $y_1 - y_2=0$. In the above formulation we define
 $g_\mathrm{hop}(x_1, x_2, y_1, y_2) := y_1 - y_2$, so that $\nabla g_\mathrm{hop}(\mathbf{x}) = (0, 0, 1, -1)$ and $ \| \nabla g_\mathrm{hop}(\mathbf{x}) \| = \sqrt{2}$. 
 Alternatively, we can integrate 
directly in the $(x,X,y,Y)$ coordinates, where
$g(x,X,y,Y) = y$, and then return to the original coordinates. This results in the following expression:
\begin{widetext}
\begin{equation}
  A_\text{hop} =
\iiiint
\limits_{\substack{x_1, x_2 = -a \\ y_1, y_2 = -b}}^{\substack{x_1, x_2 = a \\ y_1, y_2 = b}}
\rd x_1 \rd x_2 \rd y_1 \rd y_2 
 \, \indicator{ (x_1-x_2)^2 + (y_1-y_2)^2 \ge (2r)^2 } \, \delta \left(\frac{y_1-y_2}{\sqrt{2}} \right).
\end{equation}
\end{widetext}
Carrying out the integrals  (see Appendix A) gives
 \begin{equation}\label{AreaH}
 A_\text{hop}  =  16 \sqrt{2} b(a-r)^2.
\end{equation}
As before, the formula is no longer valid for $r > w/4$. In this case,
vertical hopping becomes impossible for a larger radius.

In Monte Carlo simulations, we count the proportion of successful placements of hard discs 
for which the distance 
$|y_1 - y_2|$ is within a small tolerance of $0$, but note that the factor $\sqrt{2}$ must \emph{also} be taken into account in this calculation (see Appendix B).
Results are shown in Fig.~\ref{AreaHopp01}.

The results for horizontal hopping are obtained by interchanging $a$ and $b$.

\begin{figure}[h]
\centering
\includegraphics[width=0.45\textwidth]{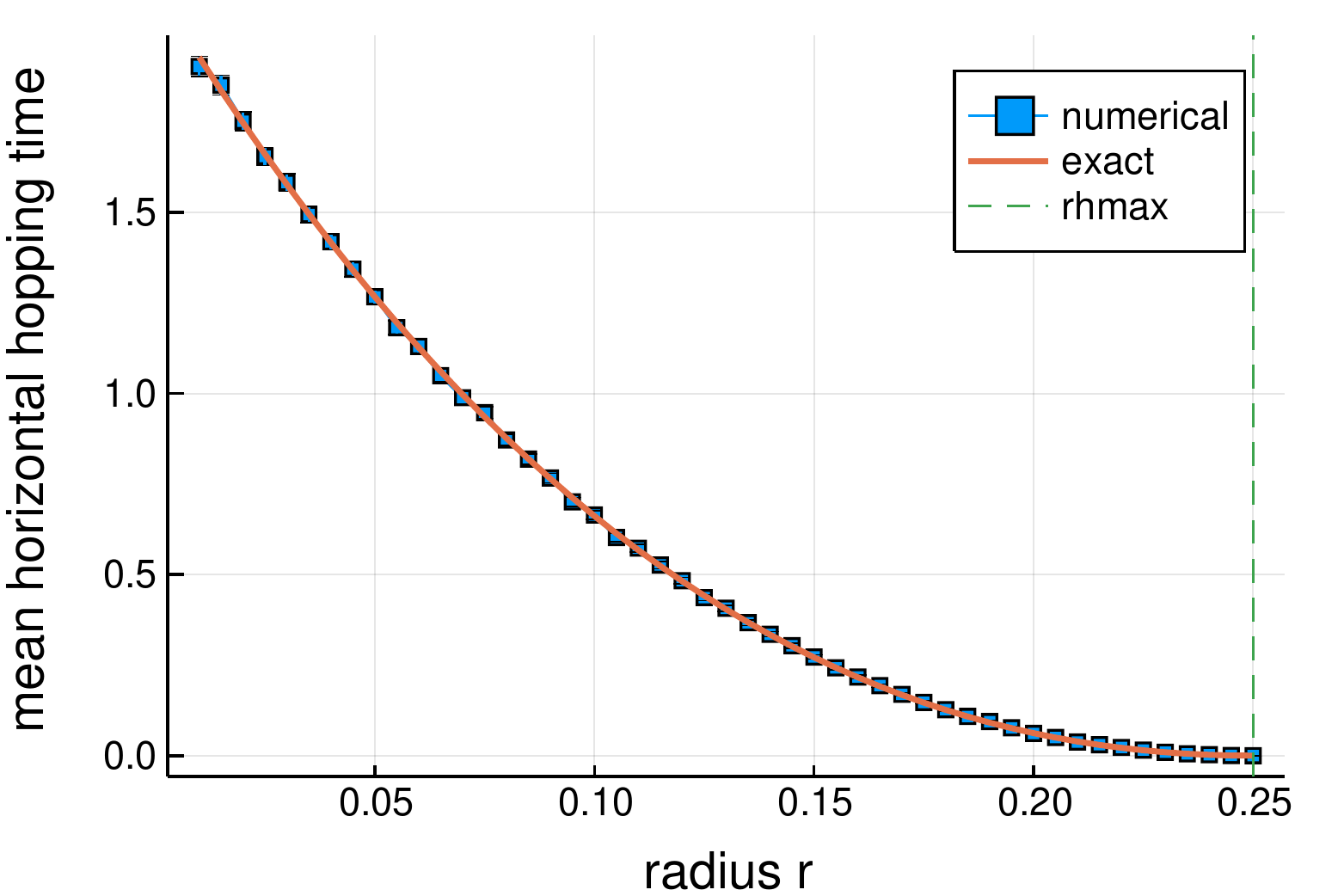}
\caption{Area of the surface corresponding to vertical, $A_\text{hop}$, 
  indicated in the formula in eq. \ref{AreaH}. As explained 
in the text, the formula does not apply for $r>1/4$, but numerical values correctly
remain at zero.}
\label{AreaHopp01}
\end{figure}

\subsubsection{Disc collisions}

The area which represents collisions between the two discs is the surface area of the cylinder
that lies within the prism, given by
$$g_\text{coll}(\mathbf{x}) := (x_1 - x_2)^2 + (y_1 - y_2)^2 - 4r^2, $$
so that
$$\nabla g_\text{coll}(\mathbf{x}) = (2 (x_1 - x_2), -2(x_1 - x_2), 2(y_1 - y_2), -2(y_1 - y_2)),$$ 
with norm $\| \nabla g_\text{coll}(\mathbf{x}) \| = 2\sqrt{2} \sqrt{(x_1 - x_2)^2 + (y_1 - y_2)^2}$.


For $r<w/4$, we find for the corresponding area
\begin{align}\label{AreaChoque}
A_\text{coll} & =  \sqrt{2} (
16\pi a b r -32 (a+b)r^2 +16 r^3).
\end{align}

\begin{figure}
\centering
\includegraphics[width=0.45\textwidth]{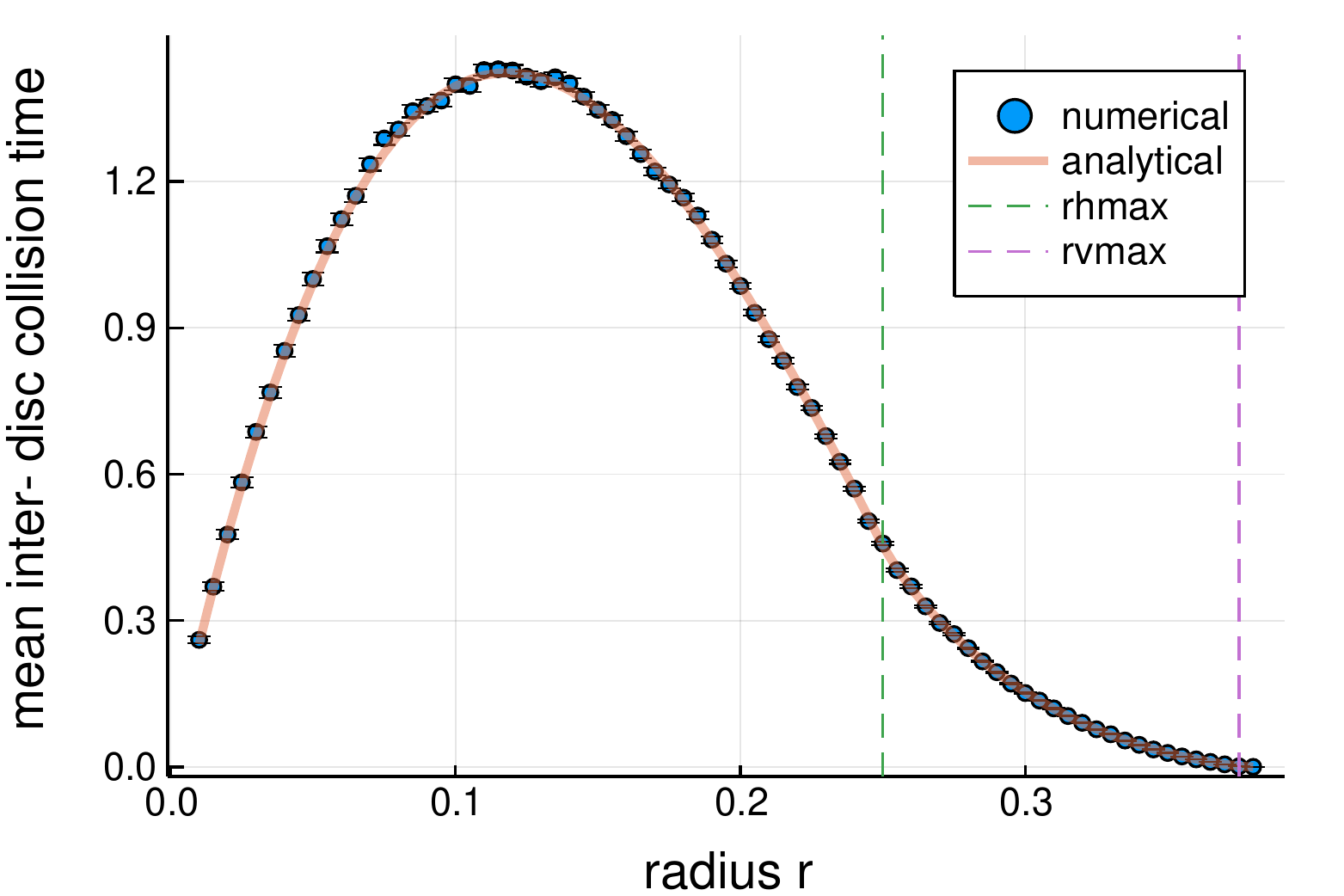}
\caption{Numerical and theoretical collision area 
between the discs.  The theoretical result
\eqref{AreaChoque} breaks down at
$r > 1/4$, but we have here used the general formula  \eqref{app:colgeneral} valid for all $r$.}
\label{AreaChoqueTeoyNum}.
\end{figure}

The numerics proceeds as before, checking which
random configurations
fall within a small tolerance from the collision condition, and
plotting this as a fraction of the total volume. The result is shown in Fig.~\ref{AreaChoqueTeoyNum}. 

\subsubsection{Wall collisions}

\begin{figure}
\centering
\includegraphics[width=0.45\textwidth]{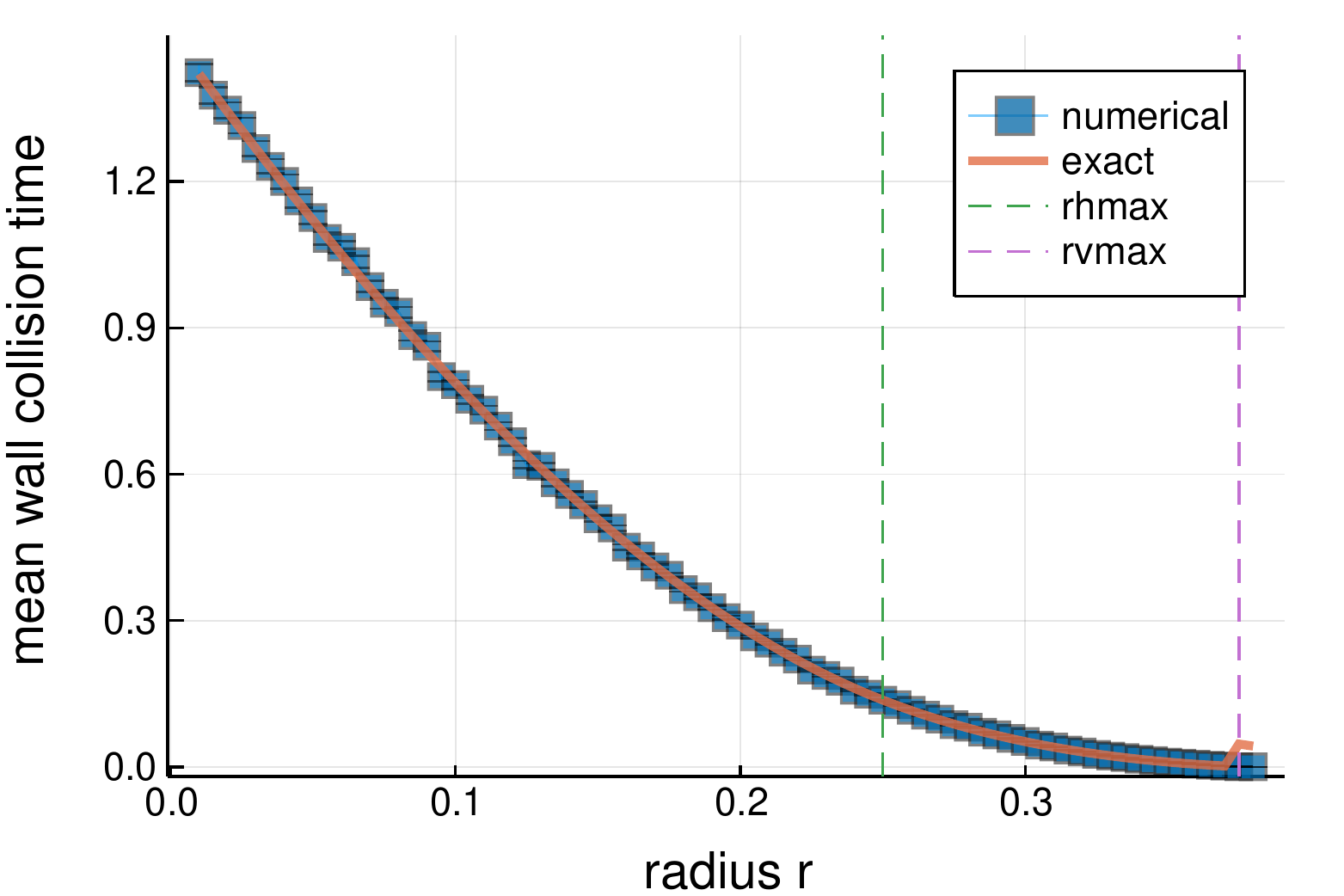}
\caption{Numerical and theoretical calculation for the area
for the impact of a particular disk with the right wall, as in \eqref{areaindic}.
Again, the general formula has been used.}
\label{area1derecha}.
\end{figure}

We restrict attention to collisions of disc $1$ with the right wall, for which
$$g_\text{wall}(\mathbf{x}) := x_1 - a.$$
The corresponding area, after taking the delta function into account, is
\begin{equation}\label{areaindic}
  A_\mathrm{wall} =  \iiint \limits_ {\substack{x_2 = -a \\ y_1, y_2 = -b }}^
  {\substack{x_2=a\\ y_1,y_2=b}} 
   \rd x_2   \rd y_1   \rd y_2 \, \indicator{ (a-x_2)^2 + (y_1-y_2)^2 \ge 4 r^2 }
\end{equation}
The integration gives
\begin{align}\label{areax1p}
 A_\mathrm{wall} & = 8 a b^2-4  \pi b r^2 +\frac{16}{3}r^3 .
\end{align}
Once again, a simple Monte Carlo procedure verifies this result,
shown in Fig.~\ref{area1derecha}. 
This time, the correction factors of $\sqrt{2}$ do not appear, since
the areas are orthogonal or parallel to the original
position coordinates.

Taking into account the symmetry of the expression for either disc
 bouncing on either of the vertical walls, the
area for this event is  $4A_\text{wall}$. For horizontal wall 
collisions, the roles of $a$ and $b$ are switched.
Thus the area for any impact on any wall is
\begin{align}\label{areawalls}
 A_\text{walls} & = 32 a b (a+b)-16 \pi r^2 (a+b) +\frac{128}{3}r^3.
\end{align}

\section{Exact mean inter-event times}

In this section, we use the results of the last section to give
analytical expressions for mean inter-hop times, as well as mean wall collision and disc collision times.
We tested the results against molecular dynamics simulations with uniform random non-overlapping initial conditions for the discs and standard billiard dynamics with elastic collisions, in which 
the trajectories of the disc centers 
are determined by their momentum and the first collision
(either with a wall or between discs) is given by
the event with the least positive collision time. 

This dynamics produces a list of collision times with walls and between discs.
Hopping times are inferred by locating consecutive collisions in which the sign of the difference of the relevant coordinates (horizontal or vertical) changes.
The exact hopping time may then be determined 
using the momenta after the previous collision.

\subsection{Mean hopping time}

Inserting the results of the previous section 
into the formula for the mean times for crossing
surfaces of section \eqref{meanfreetime} gives exact mean inter-hop times.
For vertical
hops we have
\begin{equation}\label{hoptau}
 \mean{\tau_\text{hop}} = 	
\frac{3 \pi}{4\sqrt{2}}
\frac{2 a^{2} b^{2}  - 2 \pi a b r^{2} + \textstyle \frac{a+b}{3}  (2r)^{3}  -  r^4}
{ b \sqrt{2}  ( a - r )^2}.
\end{equation}
(Recall that here there is a factor $s = 2$.)

In the limit of small disc radius, the discs have very infrequent interactions, and the result depends only
on the table height:
\begin{equation}\label{hoptaulimit}
 \mean{\tau_\text{hop}} \overset{r \to 0}{\sim}
\frac{3 \pi}{8\sqrt{2}}h.
\end{equation}

Also of interest is the limit when $r \to w/4$, at which point vertical hopping becomes
impossible.  The denominator goes to zero quadratically, while the available volume
is still positive (except in the degenerate case $w=2h$). The exact expression
is cumbersome, due to the fact that the general volume expression contains trigonometric functions
 and is unintuitive (see Section~\ref{app:volume}),
but the leading term in the numerator is  $w^2 (\frac{h}{2} - \frac{w}{4})^2$, and the denominator
stays the same, so the asymptotic behavior is $\sim (r - \frac{w}{4})^{-2}$. The figure \label{MeanHopp01} shows the comparison between
numeric and analytic results for the complete valid range. 

\begin{figure}[h]
  \centering
  \includegraphics[width=0.45\textwidth]{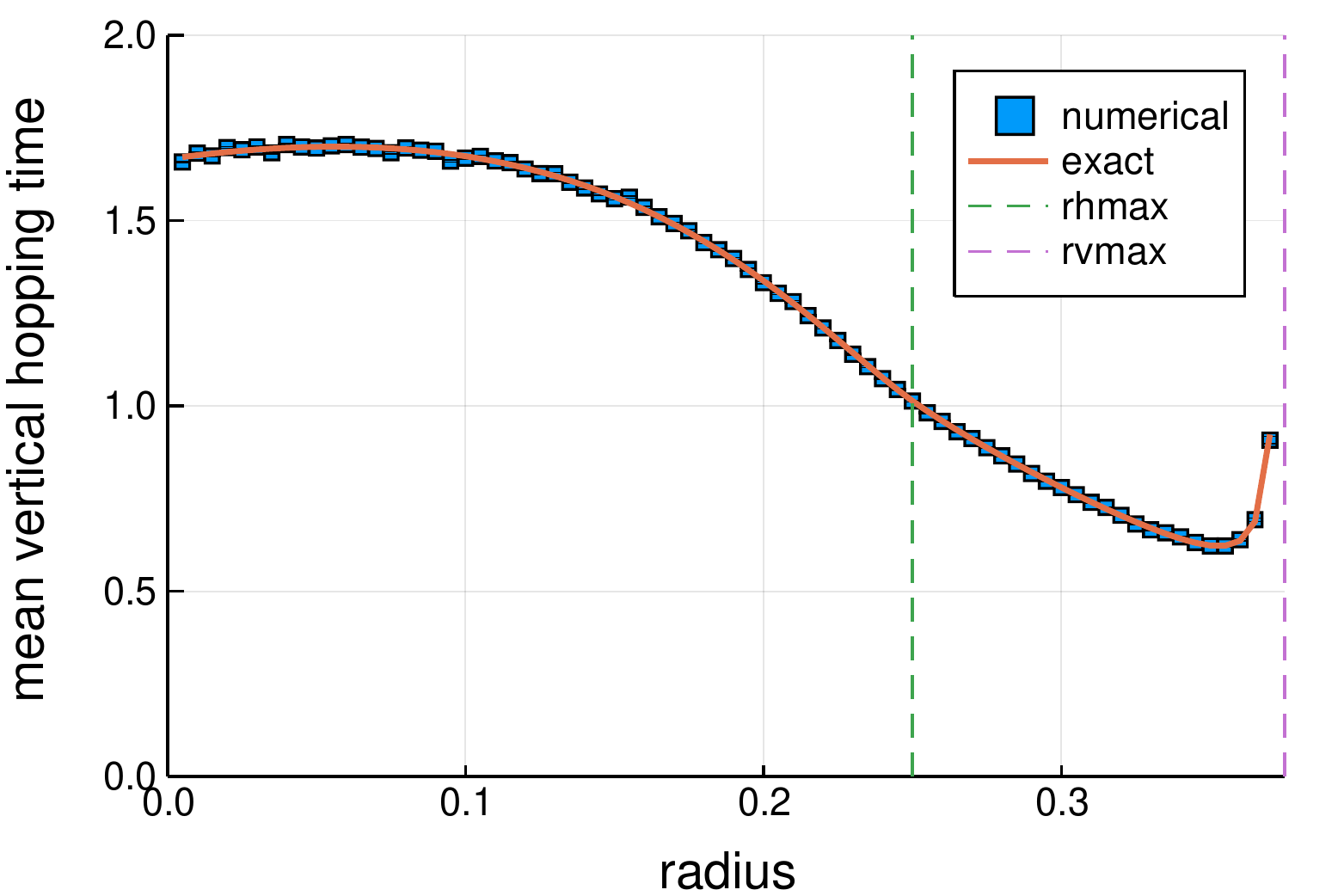}
  \caption{Mean hopping time as function of the radius.}\label{MeanHopp01}
\end{figure}

\subsection{Mean disc collision time}

For collisions between discs, we have, for the case $r < w/4$,
\begin{equation}\label{colltau}
 \mean{\tau_\text{coll}} = 	
\frac{3 \pi}{2\sqrt{2}}
\frac {2 a^{2} b^{2}  - 2 \pi a b r^{2} + \textstyle \frac{a+b}{3}  (2r)^{3}  -  r^4}
{2\pi a b r -4(a+b)r^2+2r^3}.
\end{equation}
As expected, this tends to infinity in the limit of small radius, with asymptotics
\begin{equation}\label{colltaulim0}
\mean{\tau_\text{coll}} \overset{r \to 0}{\sim}
\frac{3}{8\sqrt{2}}\frac{wh}{r}.
\end{equation}

\begin{figure}[h]
  \centering
  \includegraphics[width=0.45\textwidth]{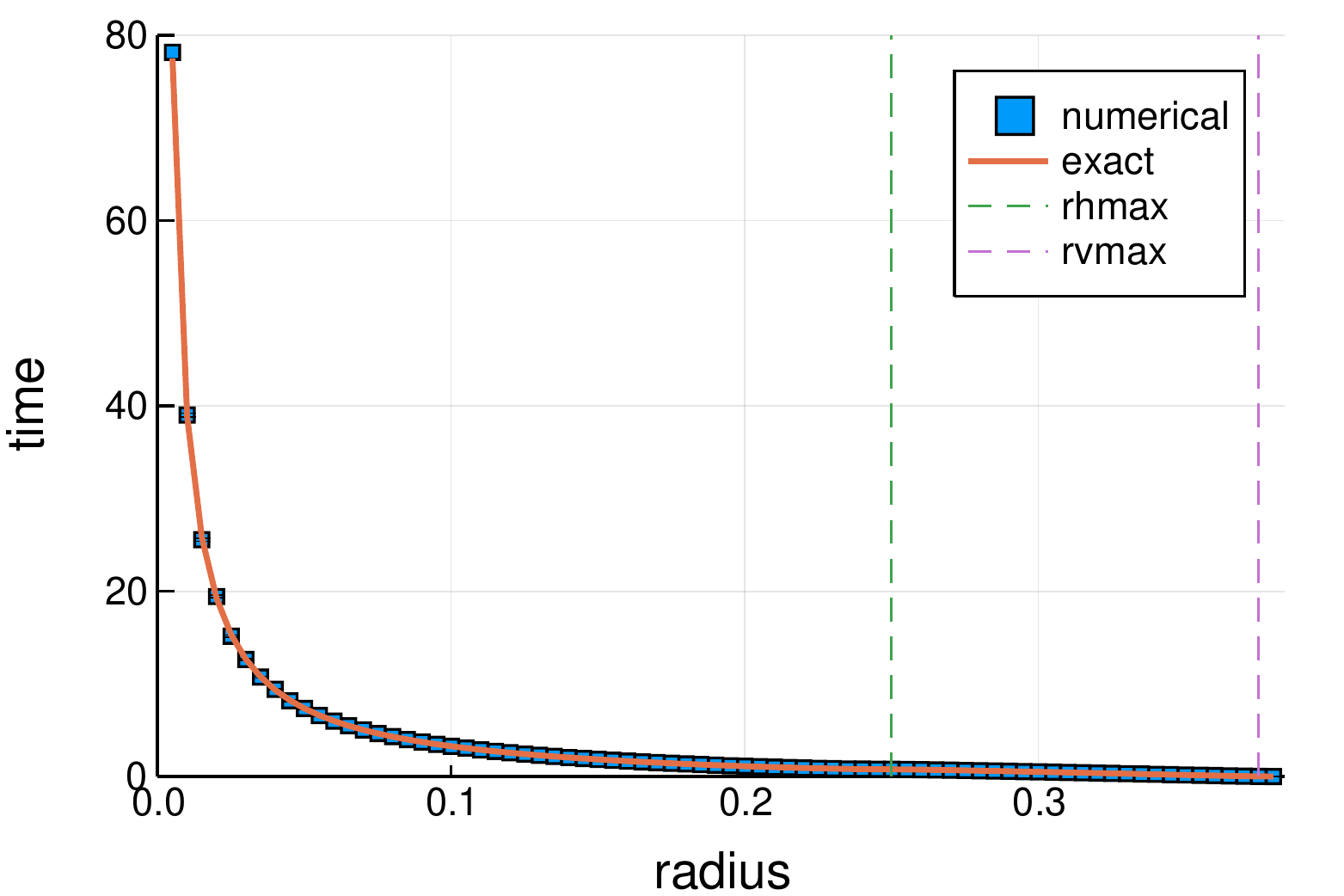}
  \caption{Mean disc collision time as function of the radius. }\label{MeanCol01}
\end{figure}

For the case in which the discs narrowly fit inside the table we need to
use the more cumbersome expression in \eqref{VolumenCasoFeo} and
the corresponding area. The time between collisions should go to zero; 
see figure~\ref{MeanCol01}. In this case, the function is smooth for the whole
range of valid values for $r$.

\subsection{Mean wall collision time}

Lastly, for a specific disc colliding with a specific vertical wall we have
\begin{equation}\label{impactwall}
 \mean{\tau_\text{wall}} = 	
\frac{3 \pi}{2\sqrt{2}}
\frac { 2a^{2} b^{2}  -  2\pi a b r^{2} + \frac{a+b}{3}(2r)^3 - r^4}
{ab^2-\pi/2b r^2 + \frac{16}{3} r^3 },
\end{equation}
in the case that $r<h/4$. The limiting form for small $r$ depends
only on the width of the table and is exactly
$3\pi w/2$. In the numerics this appears as the intersection with the
$y$-axis. 

When $r\approx h/4$ we have two limiting expressions
since there is a discontinuity in the available volume; the difference is exactly
a factor of two. In the comparison with numerics, we can see
this jump in the function, accompanied by a larger error in the
calculations (since it is more difficult to find correct starting positions). 
This is indicative of 
the system breaking up into smaller ergodic components.

\begin{figure}[h]
  \centering
  \includegraphics[width=0.45\textwidth]{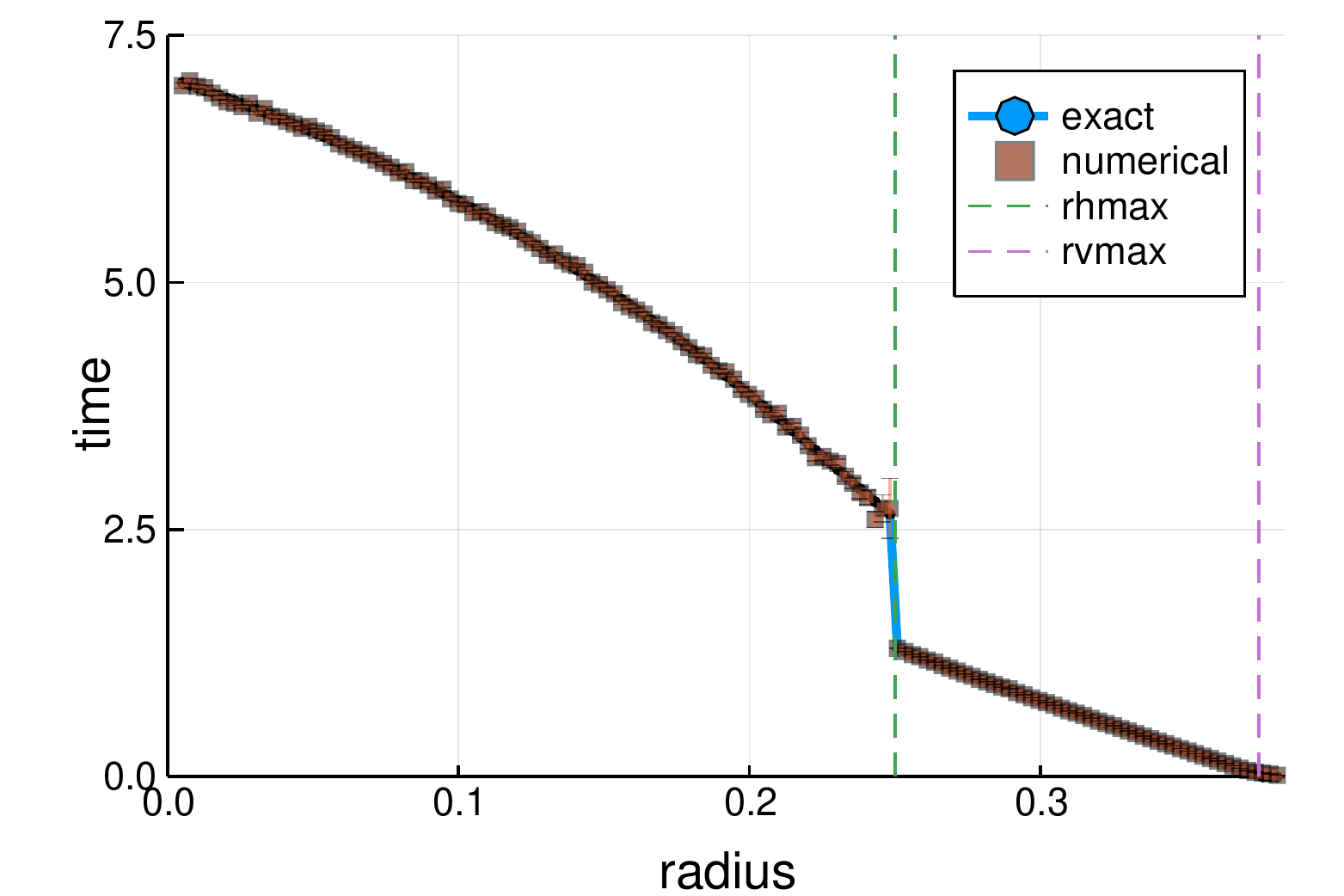}
  \caption{The mean impact-on-walls time as function of the radius, 
    notes as the figure \ref{MeanHopp01}}
    \label{MeanImp01}.
\end{figure}

\section{Probability distributions}

Since each type of event studied is a recurrence time to a
certain surface in phase space, we expect to observe the standard
exponential distribution for return times in chaotic systems
\cite{Hirata1999}. It is known that deviations from the
exponential depend on the particularities of the system at hand
\cite{Altmann2005}.


As an example, we show
the probability distribution of hopping times for different radii and find that the
distribution depends both qualitatively and quantitatively on the
radius, with the tails approximately exhibiting the expected exponential decay.

\begin{figure}[h]
  \centering
  \includegraphics[width=0.45\textwidth]{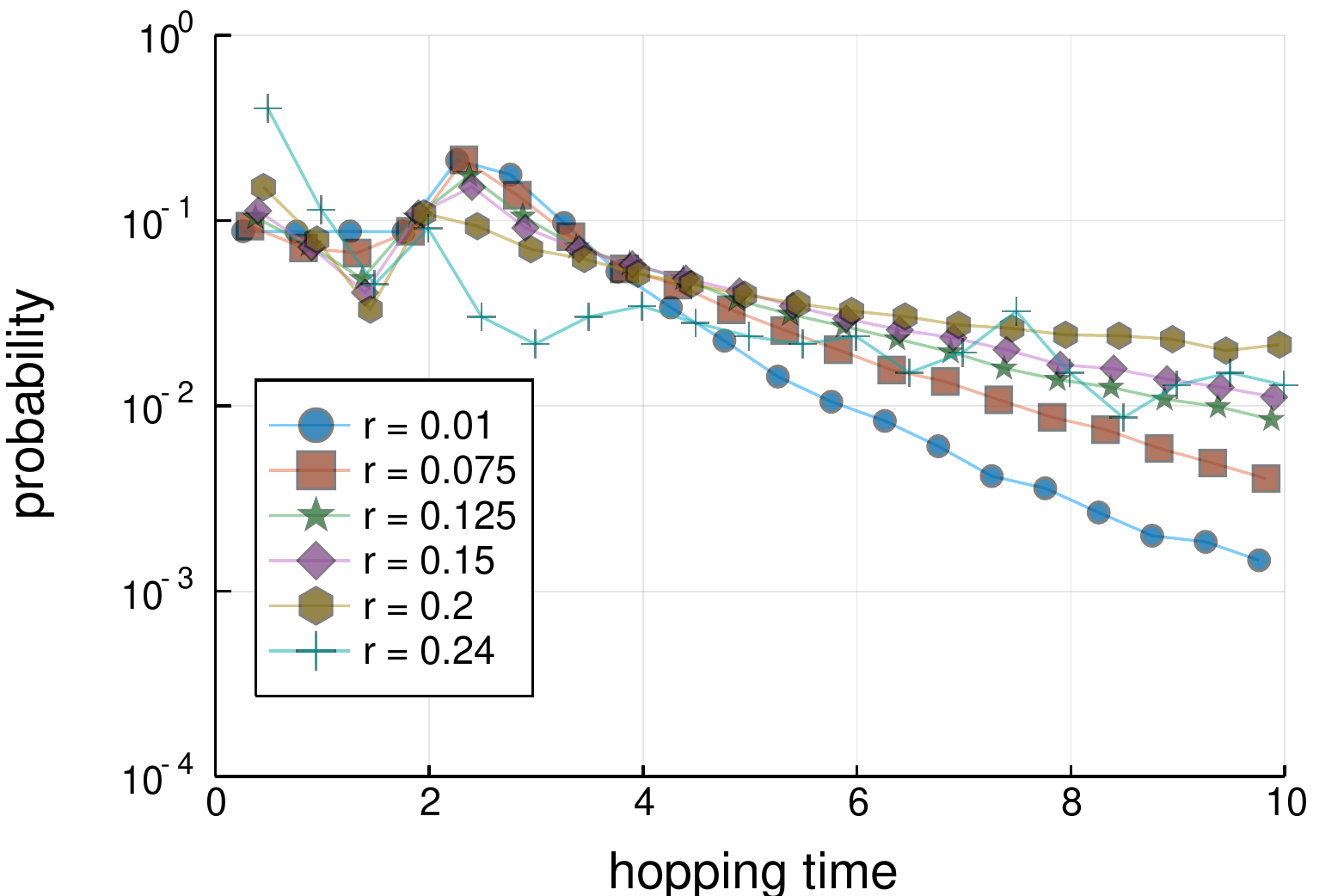}
  \caption{Histogram of hopping times for different disc radii.}
    \label{histogram_hopping}
\end{figure}

\section{Conclusions}

We have calculated exact analytic expressions for mean inter-event times for hopping and
collision in the paradigmatic model of two discs in a hard box,
by treating them as returns to suitable surfaces in configuration space, 
and applying a result from ergodic theory.
The analytical results derived 
are in agreement with  the limiting behavior obtained
by Bowles \etal~\cite{Bowles04} for first-passage events, and are in excellent agreement with molecular
dynamics simulations.

The results may be extended to discs with different radii
$r_1$ and $r_2$, leading to clumsier analytical expressions.
It would also be possible to extend the results to calculate exact mean collision
and hopping times for systems containing more discs,
but the analytical calculations become more challenging; see e.g.~\cite{three_hard_discs_2004}.

\appendix
\section{General area and volume calculations}
\label{app:area_volume}

In this section, we find expressions for areas and volumes in all 
regimes.
To simplify the formulas, we omit the dependence
of $a$ and $b$ on $r$.

\subsection{Volume}\label{app:volume}

An implicit assumption was made on the limits of integration
in \eqref{integraltotal}. If $w, h > 4r$, then the limits of integration
are unaffected by the radius of the circles.
In order to avoid the same positive term $16a^2b^2$,  
in the formulas, we work here with
the excluded volume $V_\text{cyl}$, instead of the available one $V_\text{free}$.

\begin{figure}[h]
  \centering
  \includegraphics[width=0.4\textwidth]{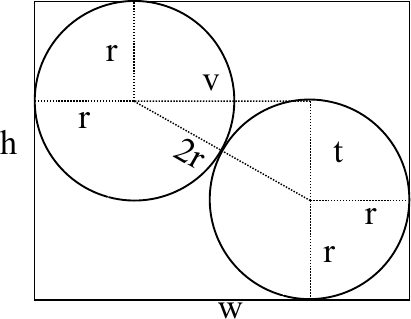}
  \caption{The largest possible radius for $h<w<2h$. The auxiliary variable
    $v$ is the horizontal difference in coordinates between the centers,
    and $t$ the corresponding vertical one.
    From the diagram
    one can see that $t^2+v^2=(2r)^2$, $h=t+2r$ and $w=v+2r$, from which
    one can deduce the value for $r$.}
  \label{radiomaximo}
\end{figure}

We begin the derivation after integrating out $X$ and $Y$:
\begin{equation}\label{VolumenGeneral}
\frac{V_{cyl}}{16}  =\iint \rd x \rd y \left[ 2ab-\sqrt{2}(ay+bx)+x y \right]
\indicator{x^2+y^2 < 2r^2 }.
\end{equation}
A diagram helps us visualize the limits of integration. The most general
case is (without loss of generality) $h < w < 2h$; as the disc radius 
\begin{itemize}
\item Horizontal and vertical hopping possible: $0 <r \leq h/4$;
\item Only vertical hopping possible: $h/4 < r \leq w/4$;
\item No hopping possible: $w/4 < r < (h+w - \sqrt{2hw}) / 2$.
\end{itemize}
The largest possible radius is illustrated in Fig.~\ref{radiomaximo}.

\begin{figure*}[h]
        \centering
        \begin{subfigure}[b]{0.32\textwidth}
          \centering
          \includegraphics[width=\textwidth]{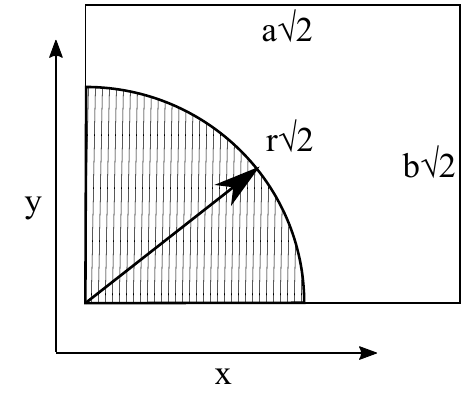}
          \caption{$2r<h/2$}
          \label{Caso1}
        \end{subfigure}%
        ~ 
        \begin{subfigure}[b]{0.32\textwidth}
          \centering
          \includegraphics[width=\textwidth]{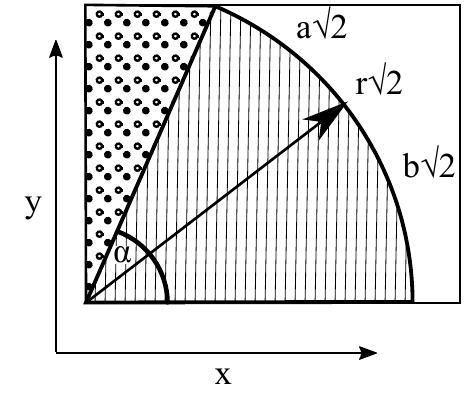}
          \caption{$h/2<2r<w/2$}
          \label{Caso2}
        \end{subfigure}%
        ~ 
        \begin{subfigure}[b]{0.32\textwidth}
          \centering
          \includegraphics[width=\textwidth]{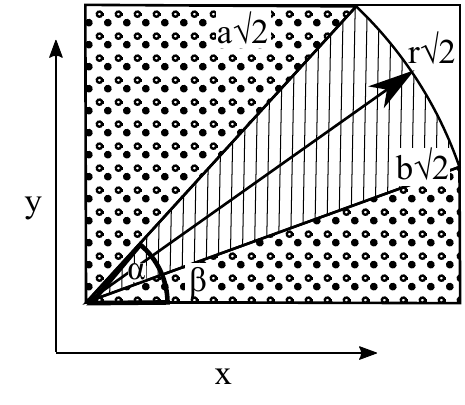}
          \caption{$w/2,h/2<2r$}
          \label{Caso3}
        \end{subfigure}%
        \caption{Three hopping regimes.
          The integral must be evaluated over the shaded region.
          It is divided in two cases: 
           the hatched area is
          solved by polar coordinates as in eq. \ref{app:volcyl},
          and the dotted area indicates the part of the integral
          that is carried away in Cartesian Coordinates as in eq.
          \ref{app:voldot}.
          In the first case all hopping is possible; the middle case
          allows only for vertical hopping; and the last case excludes hopping.}
\label{CasosIntegra}
\end{figure*}

We examine these regimes on the integration space. The first one is solved
in the main text, but we repeat it here using a different coordinate system
that proves useful. In Fig.~\ref{CasosIntegra} we present the three regimes as
the shaded area where we perform the integration. 
  
We evaluate the integral, carefully choosing the limits of integration. We perform the integral over the hatched area in polar coordinates $(\rho, \theta)$.
We call the integral $V_h(r,\alpha,\beta)$ (where $h$ stands for ``hatched''). 
As can be seen from figure \ref{CasosIntegra}, the regime where all hopping is possible 
corresponds to $\alpha = \pi/2, \beta=0$; the vertical hopping regime is where
$\alpha < \pi/2, \beta=0$, and the no-hopping regime is where $0 < \beta< \alpha < \pi/2$.

We rewrite the integral in terms of an angle $\theta$:

\begin{equation}\label{app:volcyl}
\begin{split}
\frac{V_h(r,\alpha,\beta)}{16} &=
\iint \rd x \rd y \left[ 2ab-\sqrt{2}(ay+bx)+x y \right]
\indicator{x^2 + y^2 < 2r^2}\\
&=
\iint \rd \rho \rd \theta \rho 
\left[ 2ab -\sqrt{2}(a\rho\sin\theta+b\rho\cos\theta) \right. \\
& \qquad + \left. \rho^2 \cos\theta\sin\theta \right]
\indicator{\rho^2<2r^2 }
\end{split}
\end{equation}
The indicator function determines the limits for the $\rho$ integration variable.
We set $\alpha, \beta$ as the other two limits and perform the
integral over $\rho$, which does not change in the three regimes.
\begin{widetext}
\begin{equation}
  \begin{split}
    \frac{V_h(r,\alpha,\beta)}{16} &=
    \iint\limits_{\beta,0}^{\alpha,r\sqrt{2}} \rd \rho \rd \theta \rho
    \left( 2ab -\sqrt{2}(a\rho\sin\theta+b\rho\cos\theta) 
    +\rho^2 \cos\theta\sin\theta \right)\\
 &=\int\limits_\beta^{\alpha}  \rd \theta  
\left[ 2abr^2 - r^3 4/3 (a\sin\theta+b\cos\theta)+r^4 (\cos\theta\sin\theta) \right].\\
\end{split}
\end{equation}
\end{widetext}
Now we integrate over the $\theta$ variable:
\begin{equation}\label{Volrtheta}
  \frac{V_h(r,\alpha,\beta)}{16} = 2r^2ab\theta
  +4/3r^3(a\cos\theta-b\sin\theta)
  +\frac{r^4 \sin^2\theta}{2} \Bigg\vert_\beta^\alpha.
\end{equation}
For the case in which all hopping is posible, the expression in \eqref{Volrtheta}
takes the values $\alpha=\pi/2, \beta=0$ and after multiplying by 16 both sides,
we recover the Munakata and Hu formula. For the other two cases, we use that
$\sin \alpha = b / r$ and $\cos \beta = a / r$.

Now we treat the dotted areas in the figures \ref{Caso2} and \ref{Caso3}. Since they are triangular, they
are more easily treated in Cartesian coordinates. We start with the upper
triangular region.
First, since we are inside the triangle the characteristic function
becomes a simple integration limit:

\begin{widetext}
\begin{equation} \label{app:voldot}
  \begin{split}
    V_{u}(r) /16 &=\iint \rd y \rd x [2ab-\sqrt{2}(ay+bx)+xy] \indicator{(x)^2+(y)^2<2r^2 }\\
    &=\int_0^{b\sqrt{2}}\rd y \int_{0}^{y\sqrt{r^2-b^2}/b} \rd x [2ab-\sqrt{2}(ay+bx)+xy] \\
   &=\int_0^{b\sqrt{2}}\rd y \bigl[2abx-\sqrt{2}(ayx+bx^2/2)+x^2y/2\bigr]_{0}^{y\sqrt{r^2-b^2}/b} \\
      &=\int_0^{b\sqrt{2}}\rd y
        \bigr[
          2aby\frac{\sqrt{r^2-b^2}}{b}
          -\sqrt{2}
          \bigr(
          \frac{ay^2\sqrt{r^2-b^2}}{b}
            +\frac{y^2(r^2-b^2)}{2b}
            \bigl)
           +\frac{y^3(r^2-b^2)}{2b^2}
           \bigl]\\
        &= \Bigr[ay^2\sqrt{r^2-b^2}-
          \frac{\sqrt{2}y^3}{3}
          \bigr(
          \frac{r^2-b^2+2a\sqrt{r^2-b^2}}{2b}
            \bigl)
            +\frac{y^4(r^2-b^2)}{8b^2}
            \Bigl]_0^{b\sqrt{2}}\\
          &=2ab^2\sqrt{r^2-b^2}
          -\frac{2b^2(r^2-b^2+2a\sqrt{r^2-b^2}}{3}+\frac{b^2(r^2-b^2)}{2}\\
          V_{u}(r)&=32ab^2\sqrt{r^2-b^2} -\frac{32
            b^2}{3}(r^2-b^2+2a\sqrt{r^2-b^2}) +8b^2(r^2-b^2).
  \end{split}
  \end{equation}
The lower dotted  region in Fig.~\ref{Caso3} gives a symmetric expression:
\begin{equation}
          V_{l}(r)=32a^2b\sqrt{r^2-a^2} -\frac{32
            a^2}{3}(r^2-a^2+2b\sqrt{r^2-a^2}) +8a^2(r^2-a^2).
\end{equation}
\end{widetext}

These expressions account for all volume available in the configuration space, but
cannot be used for calculation of all event times that interest us. We need also
expressions that take into account that this space is divided into disjoint components,
as some events become impossible in each of these subsystems. For example,
(see the next section), disc 1 cannot hit the left wall if
it started on the right and horizontal hopping is excluded. So we have to
take into account that only half of the positions are available (due to symmetry)
 in \eqref{meanfreetime}.

Due to the symmetry of the problem, the available volume
for each disjoint component of the dynamical system is equal; for example, if horizontal hopping is no longer possible
($w/4<r<h/4$), then there are two symmetric disjoint components: the system
starting with disc 1 on the left and the one starting with disc 1 on
the right. Both occupy the same phase space volume, as they are
symmetric under interchange of labels. If $h<4\leq r$ then the system gets further
divided into four disjoint components.

\subsection{Area}

The above procedure must be repeated for the calculation of 
areas for each surface of interest. We use a suitable Dirac delta to represent each codimension-1 collision
event (contact of a disc with a wall or another disc). We multiply the characteristic function of the available space by the Dirac delta, and
then again divide it into the three cases, namely, all hopping, only vertical hopping
and no hopping possible, again referring to Fig.~\ref{CasosIntegra}.
Sometimes it turns out to be easier to obtain 
the integral  over all configuration space and then exclude the part that
corresponds to the overlapping condition. Again, the hatched area of the exclusion
condition has a simpler representation in polar coordinates, and the triangular
(dotted) regions can be treated in rectangular coordinates.
%

\subsubsection{Hopping cross section}

\begin{figure}
\includegraphics[width=0.5\textwidth]{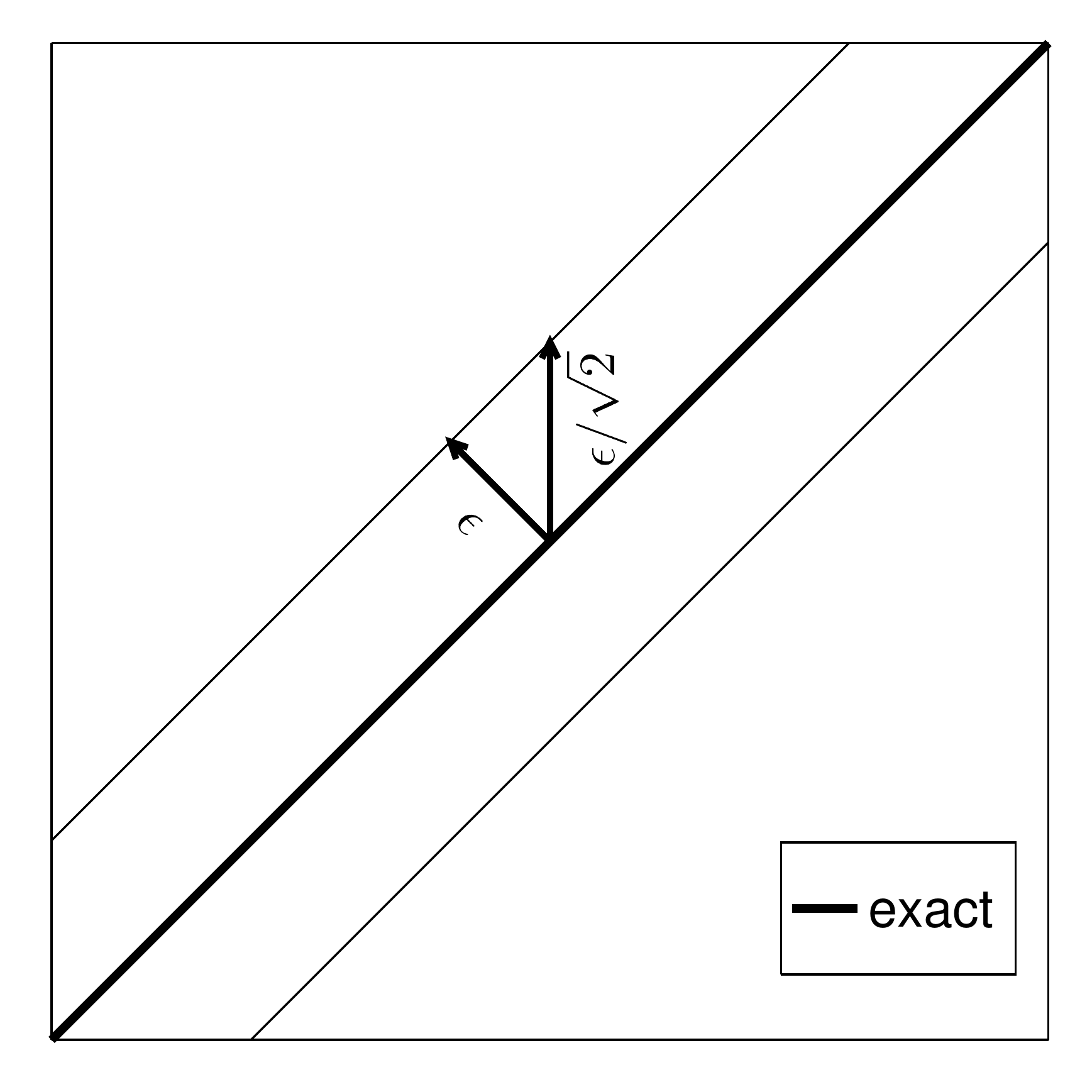}
\caption{The surface $y_1=y_2$ and the approximation to its measure by
  $\epsilon$-width characteristic functions. Using the original coordinates
  we perform an error of $\sqrt{2}$, but using the $y,Y$ coordinates
  we arrive to the correct expresion. Using the MonteCarlo Numerics, this
  error is dificult to detect, because both the numeric and the wrongly solved
  analityc formula have the same multiplicative mistake. Is only when we plug this into
Machta Zwanzig Formula that the error becomes evident. }\label{DiagramaDelta01}
\end{figure}

Fig.~\ref{DiagramaDelta01} shows the geometry for the hopping surface,
given by $g_\text{hop}(\mathbf{x}) = y_1 - y_2 = 0$. We obtain

\begin{widetext}\label{ahopcart}
\begin{equation}
  A_\text{hop} =
\iiiint
\limits_{\substack{x_1, x_2 = -a \\ y_1, y_2 = -b}}^{\substack{x_1, x_2 = a \\ y_1, y_2 = b}}
\rd x_1 \rd x_2 \rd y_1 \rd y_2 
 \, \indicator{ (x_1-x_2)^2 + (y_1-y_2)^2 \ge (2r)^2 } \, \delta \left( \frac{y_1-y_2}{\sqrt{2}} \right).
\end{equation}
\end{widetext}

The simplest approach to this integral is via the $y$ and $Y$
coordinates, since the surface is orthogonal to the $Y$ axis:
  \begin{align} 
    \frac{A_\text{hopp}}{8} & = \iiiint \limits_
      {\substack{x, X, Y=0 \\ y=b\sqrt{2}}}^
               {\substack{x=a\sqrt{2}, y=b\sqrt{2} \\
                   X=a\sqrt{2}-x, Y =b \sqrt{2}-y }}
                \mkern-36mu
     \rd x   \rd y  \rd X   \rd Y
     \indicator{x^2+y^2 \geq 2 r^2} \delta (y)
     \label{pasoraro}
     \\   
     &=  \iint \limits_{x=0, y=-b\sqrt{2}}^{x=a\sqrt{2}, y=\sqrt{2}}
    \mkern-18mu  \rd x \rd y 
    (a\sqrt{2}-x)(b\sqrt{2}-y)
    \indicator{x^2+y^2 \geq 2 r^2} \delta (y)
    \\
    &= \int \limits _0^{a\sqrt{2}} \rd x
    b\sqrt{2} (a\sqrt{2}-x)
    \indicator{x^2\geq 2 r^2}
    \\
    &= \int\limits_{r\sqrt{2}}^{a\sqrt{2}} \rd x
    (2ab-xb\sqrt{2})
    \\
    &=\biggl[2abx-\frac{x^2b}{\sqrt{2}} \biggr]_{r\sqrt{2}}^{a\sqrt{2}}\\
      A_\text{hopp}&=8\sqrt{2}b(a-r)^2
  \end{align}
  Note that in \eqref{pasoraro} we do not use symmetry in
  $y$, due to the Dirac delta.


  \subsubsection{Disc collisions}
  For other cross-section areas we proceed in a similar manner. 
  Disc collisions occur when $\sqrt{x^2 + y^2} = r \, \sqrt{2}$, giving
  \begin{equation}
    \frac{A_\text{col}}{16}  = \iiiint\limits_
         {x,X,y,Y=0}^
         {\substack{x=a\sqrt{2},\, X=a\sqrt{2}-x
             \\ y =  b\sqrt{2},\,  Y=b\sqrt{2}-y}}
         \mkern-18mu
    \rd x \rd X \rd y \rd Y
    \delta (\sqrt{x^2+y^2}-\sqrt{2}r)
  \end{equation}
  \begin{multline}
    \frac{A_\text{col}}{16}  = \iint \limits_{x,y=0}^
      {x=a\sqrt{2},\, y=b\sqrt{2}}
    \mkern-18mu \rd x \rd y 
    \bigl[ 2ab-\sqrt{2}(ay+bx)+xy \bigr] \\
    \times
    \delta (\sqrt{x^2+y^2}-\sqrt{2}r)\\
    \end{multline}
  We change to polar coordinates as we did in \ref{app:volcyl}:
  \begin{align}
    x^2+y^2 =: \rho^2 \\   \Rightarrow   \delta(\sqrt{x^2+y^2}-\sqrt{2}r) \rightarrow
    \delta(\rho-\sqrt{2}r),   
    \end{align}
    giving
  \begin{widetext}
    \begin{align}\label{app:colgeneral}
      A_\text{col}/16 & =\iint\limits_{\rho=0, \theta=\beta}^{\rho=r, \theta=\alpha}
      \mkern-18mu
    \rd \theta \rd \rho \rho
    \bigl[2ab-\sqrt{2}\rho(a\sin\theta+b\cos\theta)+\rho^2\cos\theta\sin\theta
      \bigr]
    \delta(\rho-\sqrt{2}r) \\
    &=\int_\beta^\alpha \rd \theta \sqrt{2}r
    \bigl[
      2ab-2r(a\sin\theta+b\cos\theta)+2r^2\cos\theta\sin\theta)
      \bigr] \\
    A_\text{col} & = 16\sqrt{2} \bigl( 2abr(\alpha-\beta)
    + 2r^2 [a (\cos \alpha-\cos\beta) -b (\sin\alpha -\sin\beta)]
     + r^3(\sin^2 \alpha -\sin^2\beta) \bigr).
    \end{align}
    \end{widetext}
    We have used the same trick as in the previous subsection to obtain a general expression
    that works even when hopping is not possible.
    In the case of $r<w/4$ (all hopping possible)
    we have the substitution $\alpha=\pi/2$ and $b=0$,
    obtaining the result stated previously
    in eq. \ref{AreaChoque}. It is practical to leave here the expression for 1/4 of the
    total area. The symmetry of the cases makes it so that when the phase space splits
    into disjoint components, the accessible volume and area scale in
    the same manner.
    
    \subsubsection{Wall collisions}

    For a collision with the wall, the $x_i$ give suitable coordinates. Once again, we
    have to be careful with the multiplicative factors that appear due to change
    of variables. Let us suppose that we want the area that represents contact of
    disc 2 with the right wall, so that $x_2 - a = 0$. Then
    \begin{multline}
      A_{x_2=a}  =\iiiint \limits_{-a,-a,-b,-b}^{a,a,b,b}
       \mkern-9mu \rd x_1 \rd x_2 \rd y_1 \rd y_2 
       \indicator{(x_1-x_2)^2+(y_1-y_2)^2 \geq (2 r)^2}
       \\ \times \delta (x_2-a)\\
      =\iiint\limits _{-a,-b,-b}^{a,b,b} \rd x_1  \rd y_1 \rd y_2 
      \indicator{(x_1-a)^2+(y_1-y_2)^2 \geq (2 r)^2} 
    \end{multline}
We change variables as follows:
    \begin{equation}
      \frac{y_1-y_y}{\sqrt{2}} =  y  \qquad \frac{y_1+y_y}{\sqrt{2}}=Y \qquad \frac{x_1-a}{\sqrt{2}}=x
    \end{equation}
    Integrating over $x$, we obtain
    \begin{align}\label{areachoquexy}
      A_{x_2=a} & =\sqrt{2} \mkern-18mu
      \iiint \limits
      _{-\sqrt{2}a,-\sqrt{2}b,-\sqrt{2}b+|y]}^{0,\sqrt{2}b,\sqrt{2}b-|y|}
        \mkern-18mu \rd x \rd y \rd Y 
        \indicator{(x^2+y^2 \geq 2 r^2}
        \\
        &=2\sqrt{2}\mkern-9mu
        \iint \limits_{-\sqrt{2}a,-\sqrt{2}b}^{0,\sqrt{2}b}
        \mkern-9mu
        \rd x \rd y (\sqrt{2} b - |y|)
      \indicator{(x^2+y^2 \geq 2 r^2}.
    \end{align}

    In the case that all hopping is possible there are no problems with the above derivations,
    but if vertical hopping is excluded then there are differences.
 If $r>h/4$, and disc 2 starts
    on the left, it will never be able to hit the right wall. So, actually, this
    is a subsystem of the whole  system in which our general formula needs
    to be adjusted. 
    When $h/4<r<w/4$, a disk cannot interchange left--right positions,
    but can still move across all vertical available space.  After $r\geq w/4$,
    the system gets split again into two more disjoint components.
    So, when we calculate the available volume, we use only the
    part of the volume corresponding to the set where the event can occur.
    
    Again, it is simpler to evaluate the last expression in eq. \ref{areachoquexy}
    in the excluded space and then to subtract that from the whole configuration
    space evaluation:

    \begin{widetext}
    \begin{align}
      A_{x_2=a} & =4\sqrt{2}\iint \limits_{0,0}^{\sqrt{2}a,\sqrt{2}b}
        \rd x \rd y (\sqrt{2} b - y)
        \indicator{(x^2+y^2) \geq 2 r^2}\\
     &=\underbrace{4\sqrt{2}\iint \limits _{0,0}^{\sqrt{2}a,\sqrt{2}b}
        \rd x \rd y (\sqrt{2} b - y)}_{\text{whole space}}
        -\underbrace{
          4\sqrt{2}\iint \limits_{0,0}^{\sqrt{2}a,\sqrt{2}b}
        \rd x \rd y \rd Y 
        \indicator{(x^2+y^2) < 2 r^2}(\sqrt{2} b - y)}_{\text{excluded space}}
    \end{align}
    \end{widetext}
    The integrals are again routine to evaluate: the whole space part
    gives
    \begin{equation}
      \frac{1}{4}A_{\text{whole}}=8ab^2.
    \end{equation}
    For the excluded part, we again evaluate the circular region
    in polar coordinates and the triangular regions, if they exist, in Cartesian coordinates.
    From the circular sector we get (see Fig.~\ref{CasosIntegra} and the following explanation
    for $\alpha, \beta$ variables):
    \begin{equation}
      A_{\mathrm{sector}}=8 b r^2(\alpha-\beta)+16r^3/3(\cos\alpha - \cos\beta)
    \end{equation}
    The upper triangle for $\alpha < \pi/2$ gives
    \begin{equation}
    A_{u}=\frac{8}{3}b^2\sqrt{r^2-b^2},
    \end{equation}
    and the lower triangle a symmetric expression with $a$ instead of $b$.



\section{Numerical method for calculating surface area}
    
All code used for the simulations is available, in keeping with the philosophy of open science.

For simplicity, we used traditional rejection sampling to calculate the free volume and cross-sectional areas.

The cross-sectional areas are calculated by fattening the surface $g(\mathbf{x}) = 0$ to $S_\epsilon := \{ \mathbf{x} : 0 \le g(\mathbf{x}) \le \epsilon \}$, or in the
case of the hopping cross-section $S_{2 \epsilon} := \{ \mathbf{x}: \|g(\mathbf{x}) \| \le \epsilon \}$.

Thus 
$$A = \lim_{\epsilon \to 0} \frac{ \int \mathbf{1}_{S_\epsilon}}{\epsilon}$$ 
is calculated by sampling points from the whole volume and rejecting those lying outside $S_\epsilon$.
 
As in the analytical calculation, the width of this strip must be calculated correctly by dividing by $\| \nabla g(x) \|$.

\section*{Acknowledgements}

DPS thanks Sidney Redner for posing the question that led to this work, and the Marcos Moshinsky Foundation for financial support via a fellowship.
Support is also acknowledged from grants CONACYT-Mexico CB-101246 and CB-101997, and DGAPA-UNAM PAPIIT IN117117.

\bibliography{TwoDiskBiblio}

\end{document}